\documentclass[useAMS,usenatbib]{mn2e}
\usepackage{graphicx}
\usepackage{aas_macros}

\title[mm and sub-mm atmospheric performance at Dome C]{Millimeter and sub-millimeter atmospheric performance 
  		 at Dome C combining radiosoundings and ATM synthetic spectra}
\author[S. De Gregori et al.]
{S.~De Gregori,$^1$\thanks{E-mail: simone.degregori@roma1.infn.it}
M.~De Petris,$^1$ B.~Decina,$^1$ L.~Lamagna,$^1$
J. R. ~Pardo,$^2$ B.~Petkov,$^3$
\newauthor 
C.~Tomasi,$^3$
L.~Valenziano,$^4$\\
$^{1}$Department of Physics, Sapienza University of Rome, Italy\\
$^{2}$Centro de Astrobiolog�a (CSIC/INTA), Instituto Nacional de T\'ecnica Aeroespacial, Madrid, Spain\\
$^{3}$Institute of Atmospheric Sciences and Climate, Consiglio Nazionale delle Ricerche, Bologna, Italy\\
$^{4}$Institute of Space Astrophysics and Cosmic Physics, National Institute for Astrophysics, Bologna, Italy}

\begin{document}

\date{Received 2012 May 3; accepted 2012 May 31}

\pagerange{\pageref{firstpage}--\pageref{lastpage}} \pubyear{2012}

\maketitle

\label{firstpage}

\begin{abstract}
The reliability of astronomical observations at millimeter and sub-millimeter wavelengths closely
depends on a low vertical content of water vapour as well as on  high atmospheric emission stability.
Although Concordia station at Dome C (Antarctica) enjoys good observing conditions in this atmospheric 
spectral windows, as shown by preliminary site-testing campaigns at different bands and in, not always, 
time overlapped periods, a dedicated instrument able to continuously determine atmospheric performance 
for a wide spectral range is not yet planned. 
In the absence of such measurements, in this paper we suggest a semi-empirical approach to perform an 
analysis of atmospheric transmission and emission at Dome C to compare the performance for 7 photometric 
bands ranging from 100 GHz to 2 THz.
Radiosoundings data provided by the Routine Meteorological Observations (RMO)  Research Project at Concordia 
station are corrected by temperature and humidity errors and dry biases and then employed to feed ATM 
(Atmospheric Transmission at Microwaves) code to generate synthetic spectra in the wide spectral range 
from 100 GHz to 2 THz. This approach is attempted for the 2005-2007 dataset in order to check its feasibility.
To quantify the atmospheric contribution in millimeter and sub-millimeter observations we are considering 
several photometric bands, largely explored by ground based telescopes, in which atmospheric quantities 
are integrated. The observational capabilities of this site at all the selected spectral bands are analyzed 
considering monthly averaged transmissions joined to the corresponding fluctuations.
Transmission and \textit{pwv} statistics at Dome C derived by our semi-empirical approach are consistent with 
previous works. It is evident the decreasing of the performance at high frequencies. 
We propose to introduce a new parameter to compare the quality of a site at different spectral bands, in terms of high transmission and emission stability, the Site Photometric Quality Ratio. The effect of the instrument filter bandwidth is 
involved on the estimate of the optical depth performed by the water vapour content knowledge.
\end{abstract}

\begin{keywords}
Site testing -- Atmospheric effects -- Submillimetre -- Cosmology: observations.
\end{keywords}

\setcounter{figure}{0}
\begin{figure*}
\includegraphics[width=84mm,angle=90]{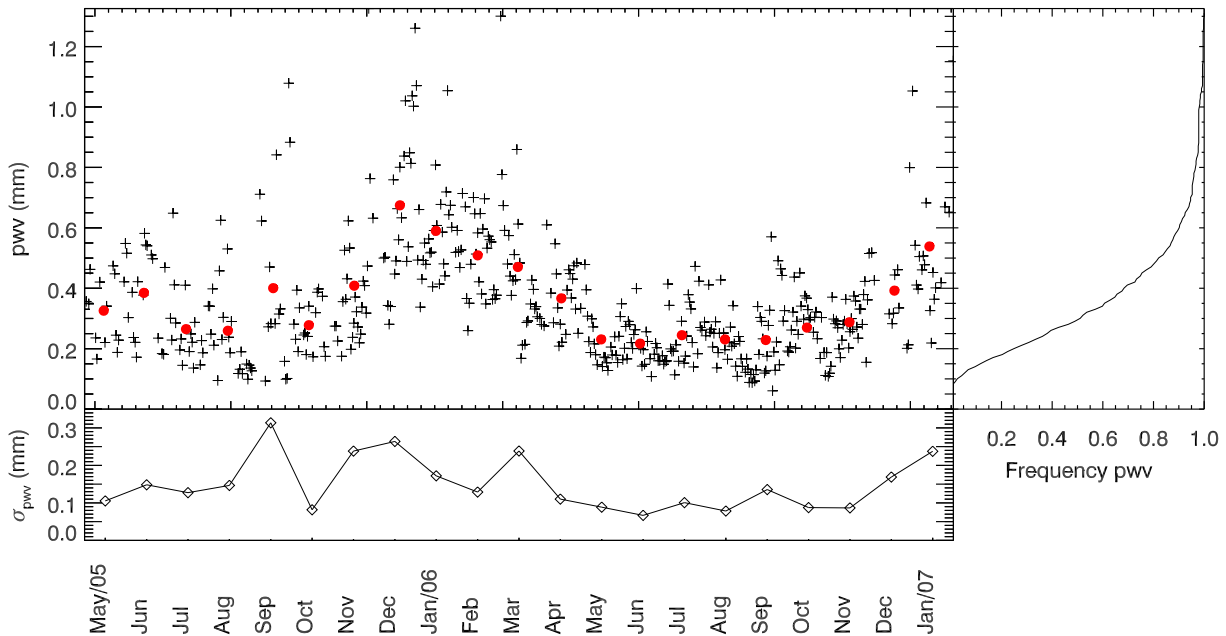}
\caption{Daily values of precipitable 
         water vapour (\textit{pwv}) estimated from the 12:00 UTC 
         radiosounding measurements performed at Dome C over the 
         period from May 2005 to January 2007 are shown in the upper left panel. 
         Monthly averages of \textit{pwv} are overplotted as red dots. 
         In the right panel the corresponding \textit{pwv} 
         vs. cumulative frequency is plotted. 
         The bottom panel shows the monthly \textit{pwv} fluctuations quantified 
         as the daily values standard deviation, $\sigma_{pwv}$.}
\label{Fig1}
\end{figure*}

\section{Introduction}  
Astronomy and astrophysics in the 100-1000 GHz band allow the study of a 
large variety of processes, in the local and distant universe, which 
involve cool matter absorbing and re-radiating efficiently at these 
frequencies, in environments often unaccessed through observations at 
visible wavelengths.
In fact, many key topics in modern astronomy and cosmology, such as galaxy 
formation and evolution, the amount and role of dark matter and dark 
energy in the universe, star formation, protoplanetary disks, or the 
properties of cold debris at the outskirts of the solar system, are 
related to radiative phenomena in this band.
The field has undergone a huge development in the last two decades, thanks 
to the development of sensitive detectors, large cameras, polarization 
sensitive devices and spectroscopically capable instrumentation. Some key 
achievements range from the measurement of the intensity and polarization 
power spectra of the cosmic microwave background at millimeter (mm) wavelengths, to 
the discovery and the characterization of the optically elusive sub-millimeter (sub-mm) 
galaxy population (SCUBA, BLAST), and the recent galaxy cluster surveys 
through sub-arcminute resolution observations of the Sunyaev-Zel'dovich 
effect (ACT, SPT).\\
Ground-based observations in the mm/sub-mm band are usually plagued by the 
transparency of the atmosphere (and its stability over time), mainly 
because of the presence of large, time-dependent pressure-broadened 
features in the emission (and absorption) spectrum of the water vapour.
Of course, this issue is strongly mitigated when operating stratospheric 
balloon-borne or airborne detectors, and completely averted when moving detectors on 
spacecrafts. BOOMERANG, BLAST, SOFIA, $Planck$ and $Herschel$ have proven the 
effectiveness of mm and sub-mm observations from the stratosphere and from 
space, providing ground breaking advancements in their respective fields 
at the time of their operation.\\
Anyway, the practical limitations on the telescope size, the weight and 
the accessibility of instrumentation still make substantially unfeasible
the deployment of large (10m class) telescopes on balloons, aircrafts or satellites. 
As a matter of fact, the ground-based solution appears presently the only 
viable way to routinely perform high angular resolution observations of 
compact objects and/or small spatial and spectral features in cool diffuse 
media at sub-mm wavelengths.\\
As a consequence, the last few years have witnessed increasing efforts in 
the design and construction of large telescopes in places of the planet
which provide the potentially most attractive atmospheric features for mm 
and sub-mm astronomy. The community has realized the need to perform a 
thorough characterization of astronomical sites in terms of atmospheric 
opacity and stability across the whole mm/sub-mm spectral region,
both for observation planning and for transparency monitoring during the 
observing sessions.\\
At a time where bolometric detectors can easily approach the photon noise 
limit, and large cameras with hundreds or thousands of such detectors
already allow to break this limitation, it is straightforward to realize 
that an improper characterization of the atmospheric properties may become 
the strongest restriction to the effective science return from 
ground-based instruments of the present (ACT, SPT) and next generation (CCAT).\\
In order to continuously monitor the atmospheric transmission
several approaches are possible: tippers or tau-meters, hygrometers, GPS, water vapour
radiometers, radiosoundings and spectrometers. 
The first approaches allow a continuous data recording by
simple instruments but with the drawback of single frequency
observations, needing a synthetic atmospheric model to infer
transmission at other frequencies.\\
Dome C is considered  one of the best sites in the world to perform observations in a wide range of the
electromagnetic spectrum allowing also to explore Terahertz windows (\cite{Minier2008} and \cite{Tremblin2011}).  
Anyway a wide frequency coverage transmission measurements campaign at Dome C, 
employing the direct spectroscopic information derived by an interferometric experiment, was never carried out.\\
The goal of this paper is to compensate the lack of those data by estimating the atmospheric 
performance with a semi-empirical approach. We test this method using the available dataset 
of radiosounding measurements recorded by the Routine Meteorological Observations (RMO)  Research Project (www.climantartide.it) 
at Concordia station in the period from May 2005 to January 2007, carefully corrected for the main lag errors and dry biases. 
The profiles of air temperature, pressure and relative humidity allow to generate synthetic spectra, ranging from 100 GHz to 2 		THz, with the ATM code (\cite{Pardo2001a}).\\
The paper is organized as follows. 
Atmospheric synthetic spectra, as derived by ATM code, are described in Sect. \ref{Synthetic}.\\
In Sect. \ref{Multi} estimates of atmospheric transmission and emission corresponding 
to largely explored ground based telescope bands between 150 and 1500 GHz 
are analyzed. The effect of the filter bandwidths on the estimate of 
opacity is for the first time included in the relation showing 
a contribution up to a 30 per cent over-estimate on the opacity in the 
case of the highest frequency band.\\ 
A parameter to rank the observational conditions for each of the selected 
spectral bands is introduced as the ratio between average transmission and 
the corresponding fluctuations.\\   
Finally a discussion on the analysis and the conclusions are summarized in Sect. \ref{conclusions}.\\
A detailed description of the correction procedure used to analyse the raw radiosounding data and determine the vertical profiles of the main thermodynamic parameters is reported in the Appendix.

\setcounter{figure}{1}  			
\begin{figure}
\includegraphics[width=84mm]{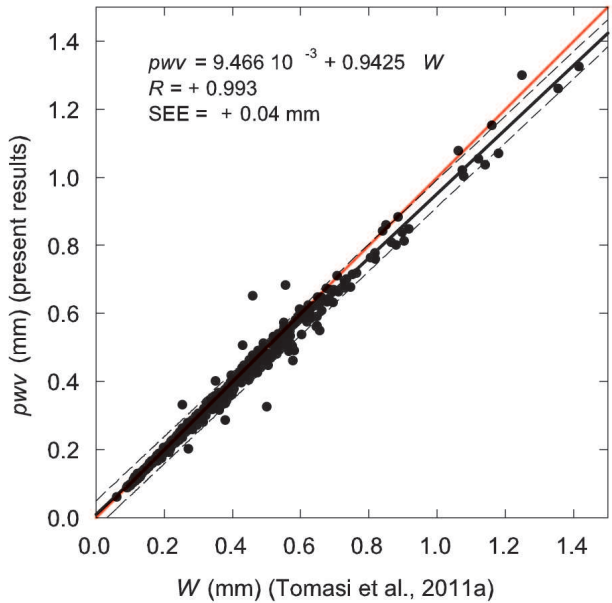}
\caption{Daily values of precipitable water vapour (\textit{pwv}) 
				 obtained through the present analysis from surface-level 
				 to 8 km amsl, and plotted versus $W$, the corresponding values 
				 of precipitable water derived  by Tomasi et al. (2011a) 
  			 over the altitude range from surface-level 
				 to 12 km amsl. The data are best-fitted by a regression 
			   line with intercept equal to $9.466 10^{-3}$ and slope coefficient 
				 equal to 0.9425, which was obtained with regression coefficient
			   R = +0.993, and provided a standard error of estimate SEE = 0.04 mm.}
\label{Fig2}
\end{figure}

\setcounter{figure}{2} 
\begin{figure}
\includegraphics[width=84mm]{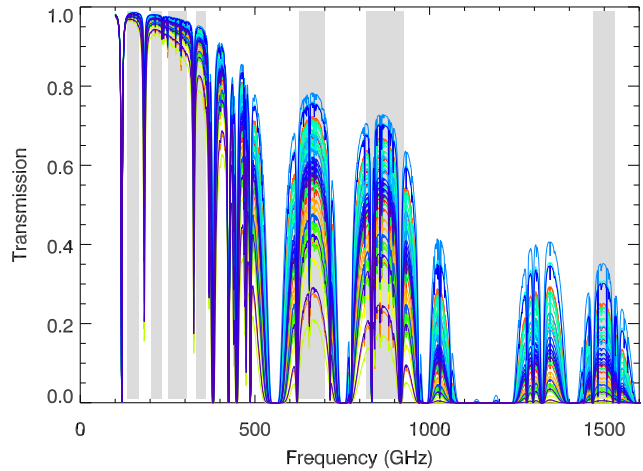}
\caption{Atmospheric transmission spectra as modeled by ATM 
         program for each radiosounding. Photometric bands in 
         Table \ref{Tab1} (gray) match the main transmission windows.}
\label{Fig3}
\end{figure}

\setcounter{table}{0}	
\begin{table}
\caption{Characteristic spectral bands assumed in this work.}             
\label{Tab1}              
\begin{tabular}{ l c c c c}     
\hline
     	                     		
  &$\nu_{0}$ (GHz) & $\lambda_{0}$ ($\mu$m) & FWHM(\%) & References     \\

 \hline
  &150     & 2000  & 22 & 1, 2, 3, 4 \\

LF&220     & 1400  & 13 & 1, 2, 3    \\

	&270     & 1100  & 18 & 1, 2, 3    \\

	&350     & 860   & 8  & 1, 2, 5, 6 \\

\hline

  &660     & 450   & 11 & 5, 6       \\

HF&870     & 350   & 13 & 5          \\

  &1500    & 200   & 5  & 7          \\              

\hline                  

\end{tabular}

\medskip
References: 
	(1) SPT,  \cite{Schaffer2011};
  (2) MITO, \cite{DePetris2002}; 
  (3) ACT,  \cite{Swetz2011}; 
  (4) BRAIN, \cite{Battistelli2012};
  (5) SCUBA, \cite{Holland1999};
  (6) SCUBA-2, \cite{Dempsey2010};
  (7) THUMPER, \cite{Ward2005}; 

\end{table}

\section{Synthetic spectra production}\label{Synthetic}	
At present, for the site of Dome C we can rely only on the atmospheric monitoring performed 
at a few individual frequencies, with no simultaneous measurements in different 
regions of the spectrum.
In order to compensate for the lack of a continuous and spectrally wide atmospheric 
monitoring at Dome C, we predict the performance in the mm/sub-mm bands in the period from 
May 2005 to January 2007 by means of a semi-empirical approach.\\
A set of raw radiosounding data was recorded for the present study, 
consisting of an overall number of 469 radiosounding measurements 
taken routinely at Dome C, at 12:00 UTC from May 2, 2005 
to January 31, 2007 ranging from a minimum of 15 in May 2005 to a maximum of 30 in November 2006.\\
In general, each radiosonde measurement consists of values of air pressure $P$, 
air temperature $T$ and relative humidity $RH$, taken at more than 800 standard 
and additional levels in the altitude range from surface to 10 km above 
mean sea level (amsl). Data provided by the radiosonde sensors are affected 
by lag and instrumental errors as well as by various dry biases. 
They were all corrected following the procedure described in the Appendix.\\
The time-patterns of the daily \textit{pwv} values are shown in Fig. \ref{Fig1}. 
Two main features are evident in Fig. \ref{Fig1}, showing that the majority of \textit{pwv} 
values are lower than 0.3 mm during the austral autumn months, although presenting largely 
dispersed patterns (of 50 per cent or more), and, hence, low stability.\\	
As shown by \cite{Tomasi2011a}, a limited contribution is given to the overall value 
of atmospheric \textit{pwv} by the amount of water vapour present in the Upper Troposphere 
and Low Stratosphere (UTLS) region from 8 to 12 km amsl, while negligible 
fractions of \textit{pwv} ranging mainly between 0.003 and 0.005 mm 
throughout the year are present in the stratosphere from 12 to 50 km
above Dome C (\cite{Tomasi2011b}).\\
To verify the reliability of the present estimates of \textit{pwv}, a comparison 
is made in Fig. \ref{Fig2} among the present daily values of \textit{pwv} and 
those correspondingly determined by \cite{Tomasi2011a} (indicated as $W$) 
using a more advanced correction procedure from surface-level to 12 km amsl. 
The comparison showed that a close relationship exists between the present results 
and those of \cite{Tomasi2011a}, defined by a regression line with nearly null 
intercept and slope coefficient of + 0.9425, having regression coefficient better 
than + 0.99, and providing a standard error of estimate equal to 0.04 mm. 
These findings clearly indicate that the present evaluations of \textit{pwv}, 
as obtained over the altitude range from surface-level to 8 km amsl, 
are fully suitable for the purposes of our study, especially considering the 
intrinsic uncertainty of the simulation model.

We have estimated synthetic spectra in emission and in opacity by means of the
ATM code in the wide spectral range from 100 GHz to 2 THz. 
Each spectrum is derived considering the corrected radiosounding data.
The transmission corresponding to each radiosounding dataset, estimated from optical 
depth spectra as $T = e^{-\tau}$, is shown in Fig. \ref{Fig3}.\\
In the period under consideration, the inferred \textit{pwv} values show an average close to 
0.3 mm with a mean dispersion of about 150 $\mu$m (see Fig. \ref{Fig1}).
The same amount of \textit{pwv} variation can contribute with a different weight to the total 
optical depth. As example in Fig. \ref{Fig4} we represent the optical depth fluctuations 
derived by ATM, quantified as the maximum dispersion, due to fluctuations of \textit{pwv} 
of the order of 150 $\mu$m around three different \textit{pwv} values (0.15, 0.5 and 1.0 mm). 
It is worthy of note that for low \textit{pwv} content, $\Delta \tau$ can be as high as 60 per cent in 
the high frequency windows.\\

\setcounter{figure}{3} 
\begin{figure}
\includegraphics[width=84mm]{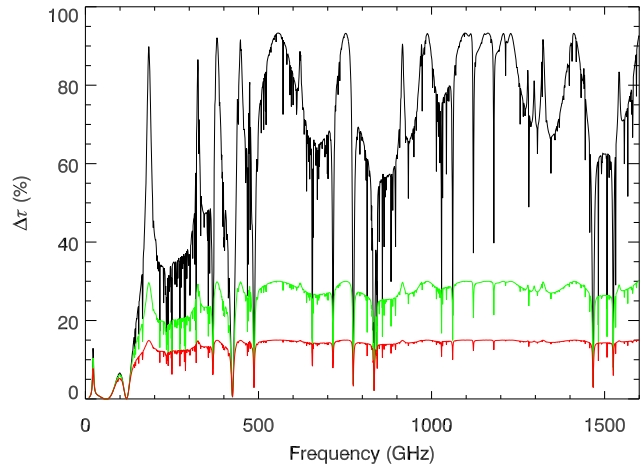}
\caption{Optical depth fluctuations corresponding to a 150 $\mu$m  variation around three selected \textit{pwv} values: 
         1 mm (red line), 500 $\mu$m (green line), 150 $\mu$m (black line).}
\label{Fig4}
\end{figure}

\section{Multi-band analysis}\label{Multi}
A quantitative analysis is performed considering 7 photometric bands 
centered at the frequencies of several astrophysical and cosmological experiments:
South Pole Telescope (SPT), Atacama Cosmology Telescope (ACT), Millimetre
and Infrared Testagrigia Observatory (MITO) and BRAIN (B-mode RAdiation INterferometer) for Low Frequency (LF) 
atmospheric windows; Submillimetre Common-User Bolometer Array (SCUBA and SCUBA-2) 
and Two HUndred Micron PhotometER (THUMPER) for sub-mm bands (High Frequency, HF). 
The central frequency of each band, as well as the bandwidth, quantified with the FWHM (Full Width Half Maximum), 
are listed in Table \ref{Tab1} (see also Fig. \ref{Fig3}). 
The band profiles are assumed to be top-hat assuming in 
this way the maximum rejection to off-band contributions.

To assess the constraints on astronomical observations arising from the atmosphere emission above Dome C, 
we give an estimate of the $NEP$ (Noise Equivalent Power) and the $NEFD$ (Noise Equivalent Flux Density) for all the
seven bands. 
In fact in such a wide spectral region both the quantities are normally employed: the power density, mainly for the low 
frequency bands, while the flux density, for the high frequency region. 
The quoted $NEP$ is the root of the sum of $NEP_{atm}^2$, the term considering the atmospheric emission fluctuations,
and $NEP_{tele}^2$, i.e. the instrumental contribution to the photon noise. 
The atmospheric emissivity spectra are generated by ATM.
The telescope is assumed a 10-m in diameter Al-mirror with 
a surface emissivity of the order 3 per cent at 150 GHz and
depending on the frequency as $\sqrt{\nu}$. 
The throughput of the telescope is assumed diffraction limited at each band.
Focal plane optical efficiencies are taken as unitary for all the bands as well as
telescope main beam efficiency.
The dominant sky sources (CMB and dust) are not  included, the instrument detector noise 
is assumed lower than the background noise and the spillover emission is neglected.
In order to quantify the maximum variation of these quantities we plot in Fig. \ref{Fig5}  
$NEP$ and $NEFD$ values for all the bands, for the extreme conditions occurred during 
the austral summers and winters at Dome C in the 2005-2007 period.

\setcounter{figure}{4} 
\begin{figure}
\includegraphics[width=84mm]{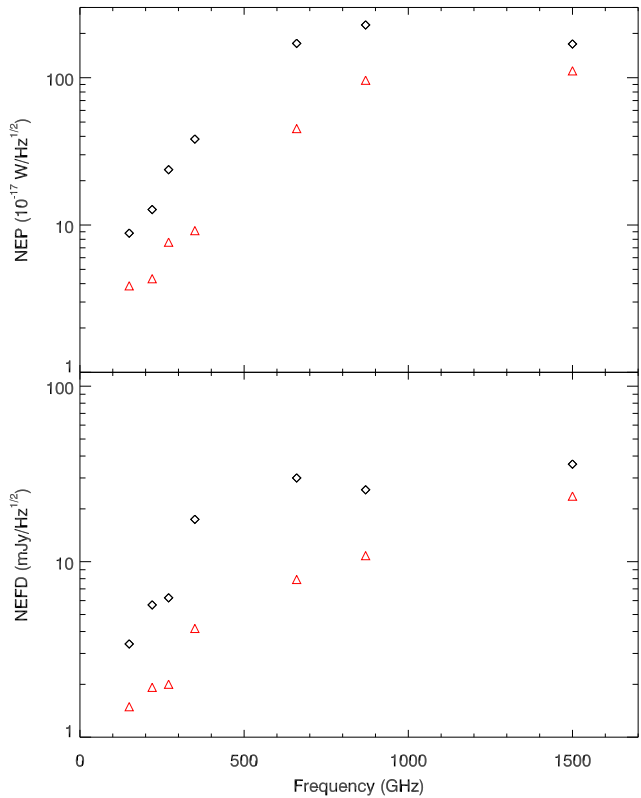}
\caption{Noise Equivalent Power (upper panel) for the seven bands in two extreme
				 atmospheric conditions in the austral winter (red triangles) and summer (black diamonds). 
				 In the lower panel, the same for the Noise Equivalent Flux Density.}
\label{Fig5}
\end{figure}

\setcounter{figure}{5} 
\begin{figure}
\includegraphics[width=84mm]{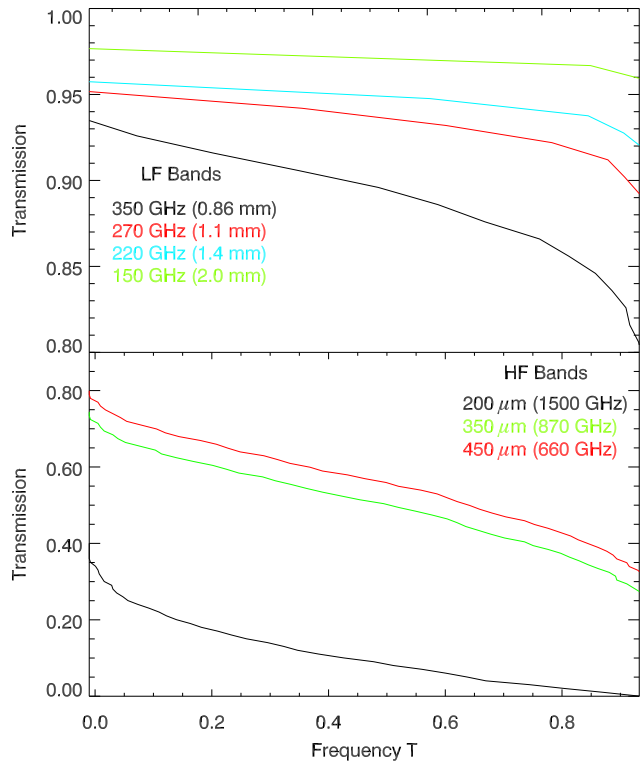}
\caption{Atmospheric transmission vs. cumulative time frequency for Dome C corresponding 
         to the atmospheric windows listed in Table \ref{Tab1}.}
\label{Fig6}
\end{figure}

\setcounter{table}{1}	
\begin{table}
\caption{Transmission quartiles matching cumulative distributions in Fig. \ref{Fig6}.}             
\label{Tab2}              
\begin{tabular}{ l c c c c c}     
\hline
     	                     		
  &$\nu_{0}$ (GHz) & $\lambda_{0}$ ($\mu$m) & 25\% & 50\% & 75\%  \\

 \hline
  &150    				 & 2000  & 0.97 & 0.97 & 0.96 \\

LF&220    				 & 1400  & 0.95 & 0.94 & 0.94 \\

	&270    				 & 1100  & 0.94 & 0.93 & 0.92 \\

	&350    				 & 860   & 0.91 & 0.89 & 0.87 \\

\hline

  &660    			 	& 450   & 0.64 & 0.56 & 0.46  \\

HF&870     				& 350   & 0.58 & 0.50 & 0.40  \\

  &1500    				& 200   & 0.15 & 0.08 & 0.03  \\              

\hline                  

\end{tabular}
\end{table}

\setcounter{table}{2}	
\begin{table}
\caption{Pwv quartiles comparison.}
\label{Tab3}
\begin{tabular}{ c c c c c c}
\hline		
Period                   & 25\% & 50\% & 75\% & References  \\				
\hline			 
01/1997                 & 0.38 & 0.52 & 0.68 & 1\\ 							  

05/2005-01/2007         & 0.20 & 0.30 & 0.45 & 2\\							

2008                    & 0.15 & 0.24 & ...  & 3\\ 							

2008-2010               & 0.21 & 0.27 & 0.35 & 4\\ 			

12/2009-01/2010					& 0.49 & 0.75 & 1.1  & 5\\		
\hline
\end{tabular}  

\medskip
References: 
(1) \cite{Valenziano1999};
(2) this work; 
(3) \cite{Yang2010}; 
(4) \cite{Tremblin2011};
(5) \cite{Battistelli2012}.
\end{table}

\subsection{Dome C statistics comparison}
To validate the proposed semi-empirical approach, 
we compare the derived atmospheric performance with the results 
available in literature.\\
Fig. \ref{Fig6} shows Dome C atmospheric transmission as a function 
of the cumulative time frequency derived by radiosounding data and ATM model for  
the bands listed in Table \ref{Tab1} (the corresponding quartiles are reported in Table \ref{Tab2}).
Transmission statistics at Dome C performed by \cite{Valenziano1999}, 
\cite{Minier2008}, \cite{Yang2010}, \cite{Tremblin2011} and \cite{Battistelli2012} 
are compared with our analysis.\\
Low frequency atmospheric windows show high transparency during the whole period 
confirming that high quality mm observations can be performed from this site for most of the time.	
For instance the 150 GHz 50 per cent quartile transmission is about 97 per cent 
(see the green line in Fig. \ref{Fig6}).  
This is consistent with the 95 per cent value recently measured by \cite{Battistelli2012} during the summer campaign 
in December 2009/January 2010, even considering their integrated in-band result.\\
Median transparency for the 220 GHz atmospheric window is about 95 per cent 
(see the cyan line in Fig. \ref{Fig6}) as already derived by \cite{Valenziano1999} 
by \textit{pwv} measurements performed with a portable photometer in January 1997.\\
Dome C 450 $\mu m$ window remains above a transmission of 60 per cent
for 50 per cent of the time. In \cite{Minier2008} the atmospheric transmission 
at Dome C has been calculated through the 5-years \textit{pwv} data from the South Pole 
available in \cite{Peterson2003} and extrapolating the corresponding atmospheric
transmission at Dome C using the model in \cite{Lawrence2004}. They found that 
450 $\mu m$ median transmission at Dome C is about 70 per cent. 
Recently \cite{Yang2010} measured a 450 $\mu m$ median winter transmission at Dome C of about 60 per cent estimating \textit{pwv} with the Microwave Humidity Sounder (MHS) sounding on the National Oceanic and Atmospheric Administration (NOAA) 
ozonesondeas in 2008.\\ 

\setcounter{figure}{6} 
\begin{figure*}
\includegraphics[width=176mm]{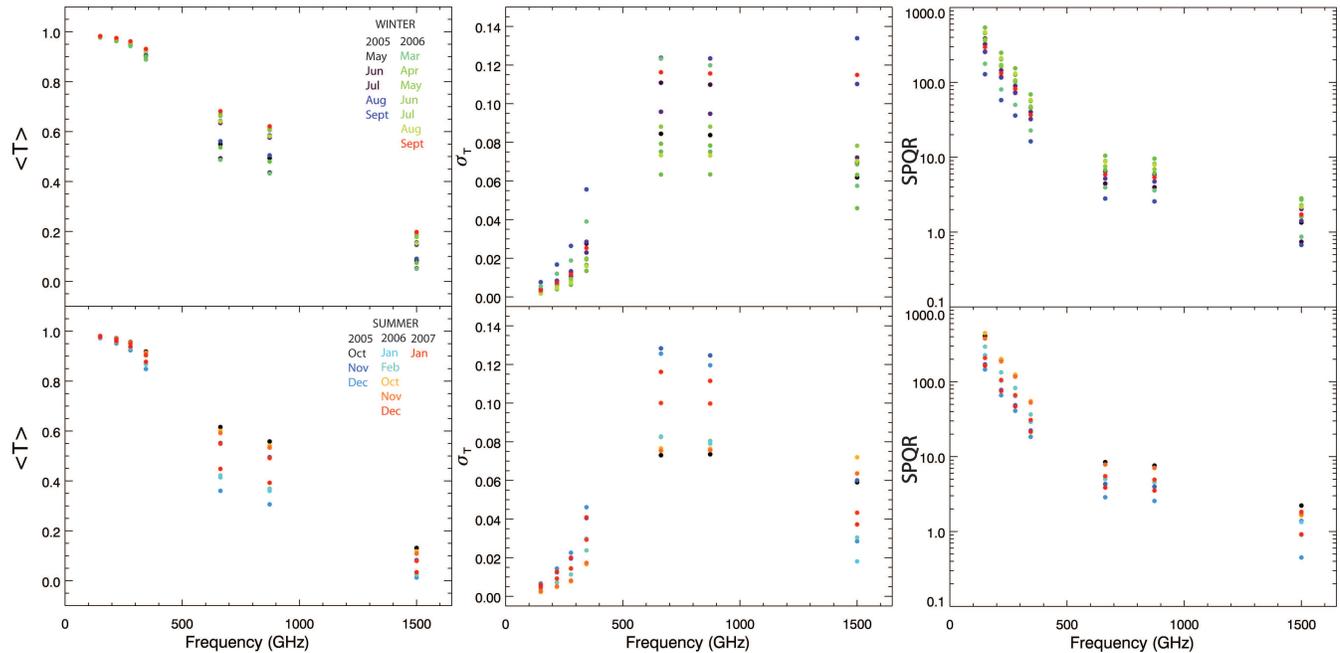}
\caption{Monthly average values of transmission $\left\langle T\right\rangle$  and the relative monthly fluctuations $\sigma_{T}$, plotted as rms values, shown in different colors: austral winter months in the upper panel and summer months in the bottom panel. Monthly Site Photometric Quality Ratio (SPQR) values are shown in the right panel. }
\label{Fig7}
\end{figure*}

Dome C median transmission for the 350 $\mu m$  atmospheric window is about 50 per cent (see the bottom panel in Fig. \ref{Fig6}), 
as derived by \cite{Tremblin2011} using the MOLIERE model and 200 $\mu m$ optical depth measurements.\\
They found also that the Dome C 200 $\mu m$ window opens with a transmission of 10 per cent 
for less then 25 per cent of the time while \cite{Yang2010} found that the transmission 
at 200 $\mu m$ is about 13 per cent for 25 per cent of the time in 2008.\\
The 200 $\mu m$ transmission as a function of the cumulative frequency
is the black solid line in the bottom panel of Fig.\ref{Fig6}: the 25 per cent quartile transmission value is above 10 per cent. \\
\textit{Pwv} quartiles since May 2005 until January 2007 (see the right panel in Fig. \ref{Fig1}) 
have been compared with Dome C water vapour estimates performed in previous works in Table \ref{Tab3}.

\subsection{High transmission and emission stability}
Following an observational approach, we report the statistics of integrated 
in-band quantities, like emission and transmission.
Both monthly averages $\left\langle T\right\rangle$  and relative dispersions 
$\sigma_{T}$ (\textit{rms} values) of in-band
transmissions, ranging from May 2005 until January 2007 and splitting 
between austral summer (from October to February) and winter months (from March to September)
are shown in Fig. \ref{Fig7}. \\

\setcounter{table}{3}	
\begin{table}
\caption{Seasonal averages of the SPQR.}             
\label{Tab4}                
\begin{tabular}{l c c c}     
\hline   	                     		
	&$\nu_{0}$ (GHz) & Summer & Winter     \\
\hline

	&150     & 272  & 335 \\

LF&220     & 127  & 152 \\

	&270     & 79  & 94 \\

	&350     & 36  & 42  \\

\hline

	&	660    & 6  	& 7 \\

HF&	870    & 5   & 6 \\

	&	1500   & 1   & 2  \\              

\hline                  

\end{tabular}
\end{table}

Monthly averaged transmission fluctuation is a good proxy of emission stability due to the fact 
that transmission and emission fluctuations are linearly correlated. 
In addition we assume that the estimated monthly averaged fluctuations, quantified in terms of the standard deviation, 
could be an underestimate of atmospheric stability because they derive from a daily data sampling, 
the time interval between two consecutive radiosoundings. \\
We note that during the austral winter the atmospheric transmission in all the
considered bands is generally higher, as expected.
$\left\langle T\right\rangle$ shows values close to the unity in mm bands and decreases towards THz windows, 
while relative dispersions $\sigma_{T}$  have the opposite trend.
As an example, the best atmospheric conditions (in term of high transmission) occur 
when the atmospheric fluctuations $\sigma_{T}$ are larger than others months  
(red dots in Fig. \ref{Fig7}). 
Referring to the atmospheric window centered at 200 $\mu$m, 
when the transmission has the maximum value, the large fluctuations at short time scales  
are likely to degrade the quality of a scientific observation.\\ 
In addition it is not possible to identify the month with the best atmospheric performance as one can see from
the gap between two consecutive years atmospheric transmission and fluctuations (red and blue dots in Fig. \ref{Fig7}). \\
450$\mu$m and 350$\mu$m bands transmission show a reduction of few percent ranging from winter to summer months,
while fluctuations are not sensitive to seasonal effects.\\
All the considered bands are characterized by high stability in October (see the black or orange dots in the middle panel of Fig. \ref{Fig7}) 
with the exception of the 200 $\mu$m window, showing high stability especially during summer months like January or February,
when atmospheric transparency is not suitable to perform astrophysical observations.\\
To quantify the real capability of the observational site we need to study 
the atmospheric performance, mainly the stability, strongly affected by the weak 
reproducibility of weather conditions at long time scales.
In order to highlight this issue we introduce a specific parameter, the Site Photometric Quality Ratio:   

\begin{equation}\label{Eq1}
SPQR=\frac{\left\langle T\right\rangle}{\sigma_{T}}
\end{equation}

relating monthly averaged transmission to its fluctuations, sampled on a daily timescale,  
for all the considered atmospheric windows. 
SPQR amplitude provides information about atmospheric performance and it allows us to dermine 
if high transmission combined with high transmission (i.e. emission) stability conditions are both satisfied for each band.
Even if we are not able to identify the desired SPQR threshold, this factor could represent 
a useful tool to compare several bands performance or sites.
In the right panel of Fig. \ref{Fig7} monthly values of the Site Photometric Quality Ratio are shown in different colors. 
The differences between the two years are more evident in SPQR, anyway a decrement of the
Site Photometric Quality Ratio towards THz regime occurs in austral winter as well as in summer periods. 
Seasonal averaged values of the  Site Photometric Quality Ratio 
in Table \ref{Tab4} suggest the good quality 
of atmospheric conditions in the low frequency bands, notably during the austral winter.
While SPQR appears useful for comparison among different bands at Dome C it 
could also be employed for comparison among different sites. It is worth reminding 
that it is difficult to quantify for SPQR a threshold value to 
discriminate the goodness of a site.\\
Two outcomes can be gathered from this analysis. If we believe in
the transmission values, as derived by this semi-empirical approach, 
a continuous atmospheric sampling is mandatory at least at high frequency to contrast the low transmission stability. 
Otherwise if we consider the data derived in this work not reliable enough for 
an accurate estimate of atmospheric properties, we need direct observational techniques.
In either case continuous atmospheric transparency measurements in all the spectral range of interest are necessary. 

\setcounter{figure}{7} 
\begin{figure}
\includegraphics[width=84mm]{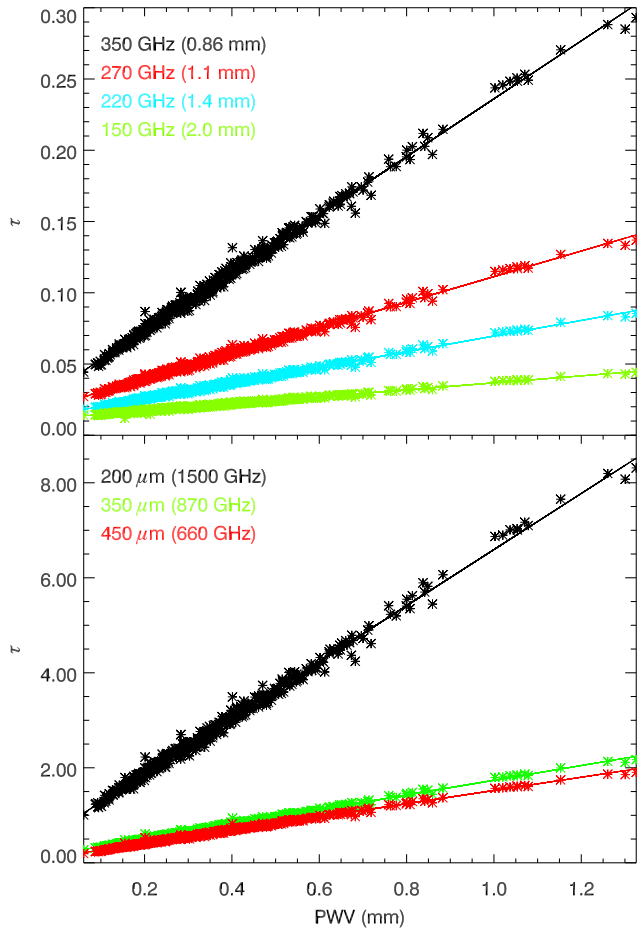}
\caption{Best fit of the correlation between the narrow-band opacity 
		   data and atmospheric water vapour for mm windows (top panel)
	   	 and for the sub-mm bands (bottom panel).}
\label{Fig8}
\end{figure}

\setcounter{table}{4}	
\begin{table*} 
\caption{Opacity-\textit{pwv} relation best fit parameters evaluated 
for the bands of interest.}
\label{Tab5}
\begin{tabular}{lccccccc}

\hline 

&	$\nu_{0}$ (GHz)       & $a_0$ & $a_1$      & $b_0$(mm$^{-1}$)  & $b_1$    & $c_0$   &  $c_1$           \\

\hline 

&150                           & 0.012 &  0.0077    & 0.024             & 0.0054   & 0.0016  & 0.0012    \\

LF	&220		  								 & 0.015 & 0.052      & 0.053             & 0.030    & 0.0024  & 0.0012     \\

&270  												 & 0.024 & 0.016      & 0.088             & 0.0070   & 0.0040  & 0.0016     \\

&360  												 & 0.032 & 0.073      & 0.19              & 0.41     & 0.0093  & 0.014      \\

\hline

&660                           & 0.16  & 0.021      & 1.35              & 1.71     & 0.066   & 0.073      \\

HF &870    										 & 0.18  & 0.68       & 1.51              & 1.96     & 0.074   & 0.070      \\

&1500                          & 0.53  & 10.13      & 5.82              & 12.69    & 0.29    & 0.35       \\

\hline

\end{tabular}
\end{table*}

\subsection{Effect of broadband filter on optical depth estimate} \label{filter}
The average of the optical depth over a band, $\tau_{\nu_0} (\Delta \nu)$, is larger than 
its central value $\tau_{\nu_0}$ so the opacity is overestimated by broadband instruments 
like tippers, as remarked as example by \cite{Calisse2004}. 
The determination of this effect is not unique because several \textit{pwv} values could give
the same in-band integrated opacity. Low-frequency instruments are less sensitive to  
this degeneracy even for large values of the bandwidth due to the flatness of the corresponding atmospheric windows. 
On the other hand a sub-mm broadband instrument overestimates the opacity (underestimates the transmission) 
and this difference depends on the filter shape as well as on the the atmospheric conditions.
Little variations of atmospheric conditions give rise to a dispersion of this overestimate
because of the relative shapes of the atmospheric window and the corresponding filter. 
For each band in Table \ref{Tab1} we have included the effect in the relation between the integrated zenith opacity
$\tau_{\nu_0} (\Delta \nu)$ and \textit{pwv} values the effect of the instrumental bandwidth $\Delta \nu$
(see Fig. \ref{Fig8} and Table \ref{Tab5}):

\begin{equation}\label{Eq2}
\tau_{\nu_0} (\Delta \nu)=(a_0 + a_1 \Delta \nu)  + (b_0 + b_1 \Delta \nu) pwv
\end{equation}

$a_0$ and $b_0$ are the linear fit coefficients of the $\tau_{\nu 0}$ vs \textit{pwv} relation 
referred to a narrow band filter matched to the central frequency and $a_1$ and $b_1$ take into 
account the dependency on the instrumental bandwidth $\Delta \nu$, 
linearly approximated at least in the range within the maximum bandwidths as reported in Table \ref{Tab1}.
Realistic band profiles could highlight the effect instead of our approximation with top-hat profiles.
The net result is that the optical depth can be overestimated at most of 30 per cent at 200 $\mu m$,
assuming the \textit{pwv} best quartile from Table \ref{Tab3}, while low frequency windows are less sensitive 
to this effect, as expected (10 per cent at 150 GHz). \\ 
The uncertainty related to the optical depth value due to the intrinsic scatter of the $\tau_0$ vs \textit{pwv} 
relation, can be approximated by a linear trend as a function of the instrumental bandwidth:

\begin{equation}
\sigma_{\tau_{\nu_0}} (\Delta \nu)=c_0 + c_1 \Delta \nu
\end{equation}

The optical depth uncertainty turns out to be 0.002 at 150 GHz and rise up to 0.3 at 200 $\mu m$,
assuming the dispersion independent on \textit{pwv} value (see Fig. \ref{Fig8}).
As a consequence the percentage uncertainty on optical depth estimate is 
about 15 per cent  all over the considered atmospheric windows assuming the best 
\textit{pwv} quartile and it remains above 10 per cent even assuming the 75 per cent quartile in Table \ref{Tab3}.\\
The six parameters corresponding to the seven bands are listed in Table \ref{Tab5}. 
The Eq. \ref{Eq2} is useful to infer the atmospheric opacity at the preferred 
frequency, with a specific bandwidth, when the \textit{pwv} content is known, but 
it is important to remind that this relation is appropriate only in the 
environs of Dome C.\\
In \cite{Tremblin2011} the opacity is related to the atmospheric \textit{pwv}
by means of the MOLIERE model. 
The resulting linear regression of the \textit{pwv} as a function 
of the 200 $\mu m$ opacity and the corresponding best fit parameters 
in Table \ref{Tab5}, neglecting $a_1$ and $b_1$, gives less than 5 per cent 
difference in transmission for low \textit{pwv} values. 
Such a gap could be easily included in the atmospheric performance variations 
observed at Dome C over the years. Also the difference in transmission evaluated for 
220 GHz best fit parameters in  Table \ref{Tab5}  and
$\tau_0(225GHz)$-\textit{pwv} linear fit in \cite{Valenziano1999} is lower then 4 per cent.

\section{Conclusions}\label{conclusions}
The quality of cosmological and astrophysical measurements performed from ground 
based observational sites in the mm and sub-mm wavelength regions 
are strongly dictated by the atmospheric performance.\\
The simultaneous measurement of atmospheric transparency and transmission fluctuations, i.e. emission stability, 
is a necessary condition to determine the true capabilities of the site of interest.\\
We try to monitor the atmosphere across a wide spectral range, mm and sub-mm, with a semi-empirical approach. 
The transmission at Dome C is inferred by generating ATM synthetic spectra as derived by radiosounding 
data in the period from May 2005 to January 2007. 
Excellent performance is evident in the low frequency bands while large emission fluctuations 
are present in the high frequency bands. In fact even if the median winter transmission is large in all 
the considered atmospheric windows, daily atmospheric emission fluctuations are not negligible and become 
remarkable in the sub-mm range.
In addition, large timescales fluctuations of the atmospheric performance have been detected during 
two consecutive years.\\
The ratio between monthly averaged transmission and the corresponding fluctuations, defined Site Photometric Quality Ratio, 
turns out to be an efficient estimator to rank the photometric performance of the atmosphere, 
in terms of stability, above Dome C, as well as any observational site.
It allows to verify when high transmission 
as well as low skynoise requirements are satisfied for the atmospheric window of interest.
The SPQR threshold for each band is not easily defined: it is depending on the detectors architecture 
and on the adopted observational strategy.\\ 
We attempted to validate the proposed semi-empirical approach comparing \textit{pwv} and transmission quartiles 
with other site-testing campaigns performed at Dome C during the last years also at different wavelengths.\\
In the usual linearly dependent opacity-\textit{pwv} relation, we include the 
effect due to the bandwidth of the monitor instrument.\\
Anyway only direct and frequent measurements of atmospheric transmission in a wide spectral range 
can provide a perfect knowledge of atmospheric influence on astronomical observations. 
If the opacity measurements are done in a narrow (a few MHz) spectral coverage, it is impossible 
to distinguish between clear sky opacity, hydrometeors contributions, and systematic errors. 
A wide frequency coverage (several hundreds of GHz) is necessary to make sure we are in clear 
sky conditions and no instrumental offset is affecting our measurement and our analysis.
In this way it is also possible to determine the dry and the wet continuum terms, see \cite{Pardo2001b}.\\
A large spectral sampling can be achieved at the price of a bit complex instrument.
The possibility to monitor the atmosphere towards different positions in the sky, also avoids 
bias due to a spatial model assuming the multi layers approximation.\\
A dedicated spectrometer, like the one proposed for Dome C (\cite{DePetris2005}) and in operation at Testa Grigia station 
(3500 m a.s.l., Alps, Italy) in a spectrally limited version (100 $\div$ 360 GHz), CASPER 2 (\cite{Decina2010} and De Petris et al. \textit{in prep.}), is a viable solution.\\

\section*{Acknowledgments}
We acknowledge Andrea Pellegrini and Paolo Grigioni for supplying us 
the radiosounding data and information obtained from IPEV/PNRA Project 
`Routine Meteorological Observation at Station Concordia -
www.climantartide.it' and Daniela Galilei for contributing to the preliminary radiosoundings data analysis.   

\appendix
\section{Correction of the radiosonde data and calculations of Precipitable Water Vapour.}
The meteorological data were obtained at Dome C using two Vaisala radiosonde models: 
(i) the RS92 model for 430 measurement days, i.e. for 94 per cent of the overall days, 
and (ii) the RS80-A model for 29 radiosonde launches only. Each triplet of signals 
giving the measurements of $P$, $T$ and $RH$ at a certain level was sent by the 
transmitter onboard the radiosonde to the ground station every 2 s. 
Considering that the radiosonde ascent rate was in general 5 - 6 m s$^{-1}$, 
the triplets of signals were recorded in altitude steps of 10 - 12 m.\\
The main characteristics of the three sensors (Barocap, Thermocap, and Humicap) 
mounted on the two radiosonde models are available in Table 1 of \cite{Tomasi2006},
where their measurement range, resolution, accuracy, repeatability in calibration, 
and reproducibility in sounding are given.\\
The measurements of $P$, $T$ and $RH$ provided by the radiosonde sensors were all corrected 
following the procedure defined by \cite{Tomasi2006}, which consists of numerous steps 
adopted to minimize the errors due to: 
\begin{enumerate}
\item the not correct calibration of the Barocap sensors;  
\item the effects caused by solar and infrared radiation heating, heat conduction and ventilation on the Thermocap sensors; 
\item lag errors, ground-check errors, and dry biases of the Humicap sensors due to basic calibration model, chemical contamination, temperature dependence and sensor aging, corrected according to \cite{Wang2002}. 
\end{enumerate}
This procedure substantially differs from that defined by \cite{Tomasi2011a} only in the parts regarding the correction of solar heating dry biases for both A- and H-Humicap sensors: (1) those of the A-Humicap sensor were corrected by \cite{Tomasi2006} using the algorithm of \cite{Turner2003}, while \cite{Tomasi2011a} preferred to use the algorithm derived more recently by \cite{Cady2008}; and (2) those of the H-Humicap sensor were corrected by \cite{Tomasi2006} using the average correction factors proposed by \cite{Miloshevich2006} as a function of solar zenith angle, while \cite{Tomasi2011a} employed the pair of day-time and night-time correction algorithms of \cite{Miloshevich2009}. A large part of the few percent discrepancies found in the comparison shown in Fig. \ref{Fig3} between the present values of \textit{pwv} and those determined by \cite{Tomasi2011a} arise from the use of these different algorithms in correcting the instrumental and solar heating dry biases affecting the field measurements of $RH$.

Using the correction procedures previously described, the daily vertical profiles 
of pressure $P(z)$, temperature $T(z)$ and $RH(z)$ were determined at fixed levels 
above the surface-level, in regular steps of 25 m from 3.25 to 4 km, 
50 m from 4 to 5 km, and 100 m from 5 to 8 km amsl.\\
In order to calculate the values of absolute humidity $q(z)$ at the same fixed levels, 
the following procedure was adopted, consisting of the three steps: 
(i) calculation at each level of the saturation vapour pressure  $E(T)$ in the pure phase 
over a plane surface of pure water, using the well-known \cite{Bolton1980} formula; 
(ii) calculation at each level of the water vapour partial pressure $e(z)$ as the 
product $E(T) \times  RH(z)$; 
(iii) calculation at each level of absolute humidity $q(z)$ (measured in $g\ m^{-3}$) 
in terms of the well-known equation of state of water vapour, and, hence, as the ratio 
between $e(z)$ (in hPa) and the product $R_w \times T(z)$ (in K), 
in which the water vapour gas constant  $R_w = 0.4615\ J\ g^{-1}\ K^{-1}$ is put in place 
of the constant $R_a = 0.287\ J\ g^{-1}\ K^{-1}$ used in the equation of state for dry air.

For all the 469 daily vertical profiles of $q(z)$ obtained using the above procedure, 
the values of \textit{pwv} were then calculated by integrating each vertical profile 
of $q(z)$ from the surface-level to 8 km amsl (i.e. up to 4.767 km 
above the ground level). 

\bibliographystyle{mn2e}
\bibliography{mybib}

\label{lastpage}
\end{document}